\documentstyle[aps,epsfig,pre,aps,epsf]{revtex}

\begin{document}
\draft
\input{psfig}
\title{Phase-separation of binary fluids in shear flow: a numerical study}
\author{F. Corberi}
\address{Istituto Nazionale per la Fisica della Materia, Unit\`a di Salerno
and Dipartimento di Fisica, Universit\'a di Salerno, 84081 Baronissi (Salerno),
Italy.}

\author{G. Gonnella}
\address{Istituto Nazionale per la  Fisica della Materia, Unit\`a di Bari
{\rm and} Dipartimento di Fisica, Universit\`a di Bari, {\rm and}
Istituto Nazionale di Fisica Nucleare, Sezione di Bari, via Amendola
173, 70126 Bari, Italy.}

\author{A. Lamura}
\address{Institut f\"{u}r Festk\"{o}rperforschung,
Forschungszentrum J\"{u}lich, 52425 J\"{u}lich, Germany}
\date{\today}
\maketitle
\begin{abstract}
The phase-separation kinetics of binary fluids in shear flow is studied
numerically in the framework of the continuum convection-diffusion equation
based on a Ginzburg-Landau free energy. Simulations are carried out for
different temperatures both in $d=2$ and in $d=3$. Our results confirm
the qualitative picture put forward by the large-$N$ limit equations
studied in \cite{noi}. In particular, the structure factor is characterized
by the presence of four peaks  whose relative oscillations
give rise to a periodic  modulation of the behavior of 
the rheological indicators and of the average domains sizes.
This peculiar  pattern of the structure factor corresponds
to the presence of domains 
with two characteristic thicknesses  whose relative abundance
changes with time.

\end{abstract}

\section{Introduction}

Binary fluids quenched below the demixing temperature exhibit an interesting
off-equilibrium phenomenology that is nowadays reasonably well 
understood \cite{bin}. Tipically, after an early stage when
domains of the two phases are formed, the kinetics proceeds by coarsening
of these structures maintaining their morphology invariant. 
For viscous fluids, such as some polymer blends, hydrodynamic effects can 
be neglected in a preasymptotic time domain and a description of the
dynamics in terms of diffusive processes alone is appropriate. 
This is the situation we focus on in this paper.
In this case,
the typical domains size $R(t)$ grows according to the Lifshitz-Slyozov
law $R(t)\sim t^\alpha\sim t^{1/3}$ \cite{lif}. 
This whole phenomenology is known to
be correctly described in terms of the time dependent 
Ginzburg-Landau model \cite{bin}. This continuum approach, beside being 
often more efficient for numerical investigations, is better suited for 
analytical purposes. In fact, despite an exact solution is presently lacking,
successful approximate theories have been developed. Among these a prominent
role is played by the large $N$ expansion. Indeed, when the model is 
generalized to a vectorial order parameter with an arbitrary number $N$ of
components, the limit $N\to \infty$ turns out to be analytically tractable
and perturbation theories around it can be applied. This special limit
has proven to provide qualitatively correct informations about the kinetic
evolution contributing, in some cases, to clarify the nature of the process
\cite{zan}. 

When segregating systems are driven mechanically, by applying an external
field or making the liquid flow, the comprehension we have
of the kinetics is much poorer. This is at odd with the
wide technological interest of these systems
 in many application areas \cite{lars}.
In the case of an applied shear flow considered here, it is known that
the evolution is profoundly changed with respect to that of a static
liquid \cite{On97}. The most noticeable effect is the alignment of the 
domains of the two species along the flow direction, chosen  in the following
as  the $x$ axis. 
This produces an anisotropy in the system and, in principle, one has to 
consider the typical domains sizes $R_x, R_y$ and $R_z$ along
the coordinate axis separately.
Modified growth exponents for  $R_x, R_y$ and $R_z$ with  respect to the 
case of an immobile fluid are expected and 
a difference  $\alpha _x - \alpha _y=1$ between the power-law exponents
of $R_x$ and $R_y$ has been reported  \cite{diff1,PT}.
In experiments, depending on the
system considered, it has also been  observed  
that, after an initial growth,  the shear 
can stabilize the system into a stationary state characterized by a very
large $R_x$ with $R_y$  finite, namely the phase separation process
is interrupted \cite{interrupt,yamam,chac1,chac2}. 

Stretching of domains requires work against surface tension.
Macroscopically 
the viscoelastic response of the fluid shows up  through 
a non-vanishing stress tensor \cite{KSH,Ro}. On microscopic scales it
has been recognized \cite{OND} that the strain exerted by the fluid
can produce different effects and in particular 
 the breakup and the recombination of domains.
The  fragmentation process is accompanied by the formation of small bubbles
and more isotropic clusters. Then these structures, which 
 grow by diffusion
and  join each other  favoured by the flow,
 are successively stretched
again and broken in a cyclical way.
In an extended system, with many such domains,
if the break-up events occur incoherently along the fluid we do not expect
any appreciable effect, apart possibly from a global slowing of the growth
process. 
Differently, if the break-up of the domains
with the successive more isotropic growth
is a coherent process
in the network occurring when the stress exceeds a sort of threshold 
reached syncronously in all the system, 
one could expect a   periodicity superimposed on the net 
behavior of the typical domains size and of other observables.

In a previous paper \cite{noi} we have shown that
a pattern similar to the latter case  outlined above 
is indeed exhibited by the large-$N$ model: Physical
observables such as the domains radii or the rheological indicators
behave for long times as power-laws decorated by log-time periodic
oscillations. In that framework it was possible to identify this
oscillatory pattern with a corresponding behavior of the four peaks
observed in the structure factor. 
 Moreover in the large-$N$ limit it
was possible to show the existence of a dynamical scaling regime 
even in the
presence of shear with  the difference between the power law growth
exponents in the flow and in the transverse directions  equal to 1.
This last observation was confirmed by a successive exact asymptotic 
solution of the large-$N$ model \cite{brayshear}. In the  
$t \rightarrow \infty$ limit,
the only accessible in this solution,  however,
 the periodic oscillations do not  survive.
Then it must believed that these are long lasting preasymptotic effects.  
In experiments, on the other hand, a four peaked structure factor has been
occasionally reported \cite{migler} while the so called double overshoot
of the excess viscosity observed sometimes \cite{double} can be interpreted
as being due to the oscillations of this quantity. 

Given this scenario the question  of the accuracy of the large-$N$ limit for
the description of real systems, where $N=1$, must be addressed.
In fact, the absence of topological defects for $N>d$ makes concepts like
domains or interfaces inappropriate and the reliability of the approximation,
even qualitatively, is not for granted. Concerning this issue important
topics to be addressed are, in our opinion, 
i) the existence of the oscillatory
pattern together with its counterpart of a four-fold peaked structure factor
and  ii) the existence of a  scaling regime with specific growth exponents.
This we do in this paper by a complete numerical investigation of the
time dependent Ginzburg-Landau model for sheared binary fluids.
A preliminary account of our results at $T=0$ in $d=2$
was already published in a previous Letter \cite{prl2}.
There the existence of structure factors with four peaks was confirmed
and interpreted in terms of domains distributed in the system
with two typical thicknesses.
Here we overview 
and detail further these results. Moreover we 
study  the effects   of thermal
fluctuations and the  role of spatial dimensionality
by comparing results for two and three dimensional systems.

Regarding issue i) our
results are confirmative of the large-$N$ prevision:  
four peaks of the structure factor are resolved and an
initial oscillation of the typical domains radii and other quantities
is observed.  Domains with different thickness and structure
factors with four peaks are observed also in quenching
at finite temperatures and in three-dimensional systems. 
With respect to point ii)
despite $R_x$ keeps growing, CPU limitations do not allow a determination 
of the the behavior of $R_y$ and $R_z$ on sufficiently long timescales to
detect a possible power-law behavior and the actual value
of the exponents. Our data only provide a rough semi-quantitative agreement
with the expected scaling laws which, therefore, cannot be inferred presently 
from numerical simulations alone.  

This paper is organized as follows.
In Sec.~(\ref{mod}) we introduce the time dependent Ginzburg-Landau 
model  used in the simulations.
In Sec.~(\ref{d2}) and~(\ref{d3}) we present the results of the numerical
simulations in $d=2$ and $d=3$ respectively. Finally
we summarize, discuss some open problem and draw our
conclusions.

\section{The model} \label{mod}

We consider the  Ginzburg-Landau free-energy 
\begin{equation}
{\cal F}\{\varphi\} = \int d^d x 
\{\frac{a}{2} \varphi^2 + \frac{b}{4} \varphi^4 
+ \frac{\kappa}{2} \mid \nabla \varphi \mid^2 \}
\label{eqn1}
\end{equation}
where $\varphi$ is the order parameter representing 
the concentration difference between the two components. 
The polynomial terms in the free-energy density
have a single-well structure when $a>0$, $b>0$ 
and describe the disordered state of the mixture with 
$\langle \varphi=0 \rangle$. 
In the ordered state, for $a<0$, $b>0$, 
two symmetric minima are located at $\varphi=\sqrt{-a/b}$. 
These are the equilibrium values of the 
order parameter in the small temperature limit. 
The gradient-squared term in
(\ref{eqn1}) with $\kappa >0$ takes into account the energy cost of
interfaces between domains of different composition.
The kinetics is described by the convection-diffusion equation
\begin{equation}
\frac {\partial \varphi} {\partial t} + \vec \nabla \cdot (\varphi \vec v) =
\Gamma \nabla^2  \frac {\delta {\cal F}}{\delta \varphi} + \eta
\label{eqn2}
\end{equation}
where the order parameter is coupled to an external velocity field $\vec v$. 
Here $\Gamma$ is a mobility coefficient and
$\vec v$ will be assumed to be a plane shear flow profile, namely
\begin{equation}
\vec v = \gamma y \vec e_x
\label{eqn4}
\end{equation}
$\gamma$ and $\vec e_x$ being a spatially uniform shear rate \cite{On97},
and a unit vector in the $x$ direction, respectively.
The thermal noise $\eta$, which 
describes thermal fluctuations \cite{On97},
has zero mean and
satisfies the fluctuation-dissipation relationship
\begin{equation}
\langle \eta(\vec r, t) \eta(\vec r', t')\rangle = 
                               -2 T \Gamma \nabla^2 \delta(\vec r - 
                               \vec r') \delta(t-t')
\label{eqn3}
\end{equation}
where $T$ is the temperature of the fluid and $\langle ... \rangle$ denotes
the ensemble average.
The validity of the present approach is restricted to systems where
hydrodynamic effects can be neglected.
For weakly sheared polymer blends with large polymerization index 
and similar mechanical 
properties of the two species, however, 
the present model is expected 
to be satisfactory in a preasymptotic time domain  when
velocity fluctuations are small \cite{bin,yeo}.

Equation (\ref{eqn2}) can be cast
in a dimensionless form after a redefinition of time, 
space and  field 
scales  \cite{par}. Then we have chosen $\Gamma=|a|=b=\kappa=1$. 

The main observable for the study of the growth kinetics is 
the structure factor 
$C(\vec k,t)= \langle \varphi(\vec k, t)\varphi(-\vec k, t)\rangle $,
namely the Fourier transform of the real-space equal time correlation
function. From the knowledge of the structure factor 
one computes the average size 
of domains in the different directions. These quantities can be
defined as
\begin{equation}
R_x(t) = \pi \frac{ \int d\vec k C(\vec k,t)}{\int d\vec k |k_x|  
C(\vec k,t)}
\label{eqnrad}
\end{equation}
and analogously for the other directions. 
Of experimental interest are also the rheological indicators,
among which the excess
viscosity defined as \cite{On97,Onuk}
\begin{equation}
\Delta \eta = -\frac{1}{\gamma}\int 
\frac {d\vec k}{(2\pi)^d} k_x k_y C(\vec k,t) .
\label{eqn5bb}
\end{equation}

We have simulated equation (\ref{eqn2}) in $d=2$ and $d=3$ by a first-order
Euler discretization scheme. 
Periodic boundary conditions have been implemented
in the $x$ and $z$ directions (for the $d=3$ case); Lees-Edwards 
boundary conditions \cite{LE} were used in the $y$ direction. These  
boundary conditions, originally developed for molecular dynamic simulation 
of fluids in shear, require the identification of a 
point at $(x,0,z)$ with the one located at $(x+\gamma L \Delta t,L,z)$,
where $L$ is the size of the lattice (the same in all the directions) and
$\Delta t$ is the time discretization interval. 
The system was initialized
in a high temperature disordered state and the evolution
was studied with $a <0$. Simulations were run using lattices of size 
$L=1024$ in $d=2$ and $L=256$ in $d=3$ with $\Delta x=0.5, 1$. 
 The width of interfaces is given by $\sqrt{2 \kappa}$ \cite{RW}. 
We do not observe significative differences between these two choices
of $\Delta x$. 
The results here
shown were obtained with $\gamma=0.0488$, $\Delta x=1$, $\Delta t=0.01$,
$\langle \varphi \rangle=0$. The two dimensional system has been studied 
considering temperatures in the range $0\leq T \leq 5$, 
while only the $T=0$ case has been studied in $d=3$.

\section{Numerical simulations: d=2} \label{d2}

\subsection{T=0} \label{T0}

In this section we discuss the results of the numerical simulations
of Eq.~(\ref{eqn2}) without the thermal noise $\eta$.
A sequence of configurations of the order parameter are shown in 
Fig.~(\ref{T0_1}) at different values of the strain $\gamma t$.
After the usual early stage, when domains are forming from the
mixed initial state, a bicontinuous structure is observed.
The distortions produced by the flow appear evident from $ \gamma t \simeq 1$
onwards. At $\gamma t=11$ striped domains aligned with a tilt angle
$\theta$ with respect to the flow are observed. As time elapses
$\theta$ decreases and the structures align with the flow.
In the meanwhile nonuniformities are formed in the system:
Regions with domains of different thickness can be clearly observed.
Small bubbles are also present, originated from the fragmentation
of strained domains. 

A more accurate analysis of the spatial properties
can be achieved by means of the structure factor, shown in Fig.~(\ref{T0_2}).
Initially $C(\vec k,t)$ exhibits an almost circular shape, corresponding
to the early stage. Then the presence of shear deforms
$C(\vec k,t)$ into an ellipse and modifies the profile of the edge until,
for $\gamma t \geq 1$, four peaks can be clearly observed. 
The presence of a peak in
the structure factor is generally interpreted as the signature of a
characteristic length in the fluid. Here, due to the anisotropy, for
each peak one associates one length for each spatial dimension.
Since the peaks are related by the symmetry $\vec k\to -\vec k$ one 
concludes that there are two distinct characteristic lengths for each
space direction. This corresponds to the observation of thinner and thicker
domains in Fig.~(\ref{T0_1}). 
The relative height of the peaks in one
of the two specular foils in which $C(\vec k,t)$ results to be 
separated starting
from $\gamma t\simeq 4$, is better seen in Fig.~(\ref{T0_3}). For a better
view of $C(\vec k,t)$ we have enlarged differently the scales on the $k_x$ and
$k_y$ axes.
Here one
observes that the two peaks dominate alternatively at the times
$\gamma t=11$ and $\gamma t=20$. In a situation like that at $\gamma t=11$,
the peak with the larger $k_y$ dominates, indicating that the stretched
thin domains are more abundant. 
When the strain is further increased
their rupture makes the 
contribution of the thick regions more important;
this causes the other peak of $C(\vec k,t)$ to dominate.
This mechanism is very reminiscent of what happens in the large-$N$
model, where the recurrent prevalence of the peaks is shown to
reproduce periodically in time up to much longer times then those
reachable in the simulations here presented.

The mechanism of stretching and break-up which characterizes 
the domains evolution produces an oscillatory pattern in the
typical size of ordered regions $R_x$ and $R_y$. This is shown
in the first picture of
Fig.~(\ref{T0_4}) \cite{notarella}. 
$R_y$ reaches a local maximum when $C(\vec k,t)$
is of the form of Fig.~(\ref{T0_3}) at $\gamma t=20$. As already
observed thick domains are more abundant at this time, as one can see in 
Fig.~(\ref{T0_1}) at $\gamma t=20$.
In the large-$N$ model
this oscillatory behavior is periodic in the logarithm of time and is
superimposed on an average power-law increase of the length $R_x,R_y$.
Here CPU limitations prevent a clear check of both these predictions.
This behavior reflects itself on most physical observables, among
which the excess viscosity. This is shown in the first picture
of the panel of Fig.~(\ref{T0_5}) where
$\Delta \eta $ is plotted against the shear strain $\gamma t$.
Starting from zero in 
correspondence of the
isotropic initial condition, $\Delta \eta$ shows a net 
increase up to a global maximum at $\gamma t\simeq 8$. This is because 
the stretching of the domains caused by the shear require work against 
surface tension. This increase, however, is not monotonous and $\Delta \eta$
shows an oscillating pattern with a first pronounced local maximum
at $\gamma t\simeq 1.5$. 
For larger times $\Delta \eta $ decreases due to the 
ordering process in the fluid.
 Fig.~(\ref{T0_5}) suggests that oscillations
decorates the behavior of the excess viscosity even in this
regime. 
 Another oscillation is completed 
at $\gamma t\simeq 40$.
 This is further supported by the analogous feature in the large-$N$
model. We mention that 
a double overshoot of $\Delta \eta$ is observed in experiments 
too \cite{double}.

\subsection{Role of temperature fluctuations}

We have performed numerical simulations of Eq.~(\ref{eqn2}) for several
temperatures in the range $0-5$. Generally speaking we observe
a similar qualitative behavior as in the zero temperature case.
In particular from the analysis of the field configurations
presented in Fig.~(\ref{Tf_1}) for $T=5$ one again observes the anisotropic
deformation of the bicontinuous pattern emerging after the linear regime
and the inhomogeneities in the domains thickness. Thermal fluctuations
are responsible for the increased roughness of the interfaces with
respect to $T=0$ and for the thermal excitations inside the ordered domains,
similarly to the case without shear. 

The structure
factor, shown in Fig.~(\ref{Tf_2}), 
starting from $\gamma t=4$ exhibits four peaks whose
relative heights prevail recurrently. However, the difference of the 
heights of the peaks 
on the same foil is less than in the case $T=0$.
At  $\gamma t=1$ instead, one has only two broad maxima along the
diagonal $k_x\simeq k_y$ that is splitted into four peaks only at later
times. The same behavior is also observed at $T=0$ but for earlier times.
This suggests that thermal fluctuations slow the evolution of $C(\vec k,t)$.

Quantitative differences with the case $T=0$, possibly related to 
the temperature-induced slower evolution,
are exhibited 
by the characteristic sizes of the domains in the $x$ and $y$ directions.  
These quantities are compared for several temperatures in Fig.~(\ref{T0_4}).
On the basis of renormalization group \cite{prl2} or scaling arguments 
\cite{brayshear} $R_y$ and $R_x$ 
are expected to obey an asymptotic
power law growth $t^\alpha$ with $\alpha_y=1/3$ and $\alpha_x=\alpha_y+1$.
The exponent $\alpha_y$ is the same as without shear and is therefore
presumably caused by the evaporation-condensation mechanism
responsible for growth in the absence of flow. On the other hand
the growth law in the flow direction is induced by the shear.
An accurate determination of these exponents
is not possible on the basis of our numerical simulations. 
Finite-size effects become quite soon relevant in the phase
separation process. Moreover the possible
 power-law behaviors are decorated by oscillations and it would
be necessary to access much longer times in order to average over this
periodic contribution. Due to finite size effects this, in turn, would
require even larger lattices, which is too numerically
demanding for standard workstations. 
Despite these limitations, by looking at Fig.~(\ref{T0_4})
one can still make some qualitative observation.
At $T=0$ one observes for $\gamma t > 30$ a regime roughly consistent with
$\alpha _x - \alpha_y =1$ even if 
$\alpha _x$ is  smaller then the expected value $4/3$ and 
$\alpha_y$ is not even well defined because
of large oscillations. By raising the temperature one sees that 
this regime is gradually changed and the effective exponent $\alpha _x$ is
lowered until, at $T=5$, it becomes comparable to the value $1/3$ 
observed without shear. If one believes that the asymptotic exponents
are not changed by the strength of thermal fluctuations, as suggested
by the renormalization group argument \cite{prl2} above mentioned
and by the analogy with the case $\gamma =0$, then the role of the
temperature is that of producing a preasymptotic behavior characterized by 
a slower growth in the direction of the flow. 

In binary mixtures  without shear 
the temperature is known to help the segregation process 
increasing the value of the amplitude in the expression 
of  $R(t) \sim  A t^{1/3}$ \cite{bin,par}.
Here, as an  effect of the increased temperature,
the contribution to the phase separation of  
the usual 
evaporation-condensation mechanism,  associated to the power exponent $1/3$,
becomes more relevant than  
the effects of the shear induced mechanism.
 This  results 
in the observed  preasymptotic behaviour  $R_x=At^{1/3}$.
At later times, even if not observed in our
simulations, we expect a growth with a larger exponent close 
to  4/3.

The above trend  is reflected in the behavior of the excess viscosity shown in
Fig.(\ref{T0_5}). 
If scaling is obeyed asymptotically, so that
$C(\vec k,t)\sim R_xR_y f(k_xR_x,k_yR_y)$, then, from Eq.~(\ref{eqn5bb})
one would expect $\Delta \eta $ to scale as the inverse domain volume,
namely $\Delta \eta \sim R_x^{-1}R_y^{-1}$. Hence, given the above discussed
behavior of $R_x,R_y$ as the temperature is changed, one expects 
$\Delta \eta$ to gradually cross-over from a power law $t^{-5/3}$
at low temperatures to $t^{-2/3}$ at higher temperatures, which
is  rougly consistent with our data. 

Finally we make some observations about the influence of the  temperature
on the oscillating behavior. 
The role of the oscillations is progressively suppressed
increasing $T$. This is observed in Fig.~(\ref{T0_5}). The most
apparent effect is the disapparence for $T>1$ of the initial
double overshoot of $\Delta \eta$. 
A gradual suppression of oscillations
due to thermal fluctuations is also observed in the plot of $R_x$ and $R_y$,
Fig.~(\ref{T0_4}).  This is consistent with the above discussion 
about the effects of raising the temperature.

\section{Numerical simulations: d=3} \label{d3}

In this section we present the results of numerical simulations of 
the zero temperature quench process in a three dimensional system.
The typical structures produced in the coarsening process  
 are shown in Fig.~(\ref{3d_1}). Domains appear with an 
interconnected cylindrical shape aligned with a tilt angle $\theta$ that
shrinks in time with the flow direction. 
Contour plots of the structure factor along the coordinate planes
are presented in Fig.~(\ref{3d_2}). On the plane
$k_x=0$ the shear term drops out from Eq.~(\ref{eqn2}) as it  can be
easily checked by transforming   Eq.~(\ref{eqn2})
 into reciprocal space. Then shear
is ineffective along this direction and the structure factor displays
the usual circular form as for a static fluid. In the plane $k_y=0$
$C(\vec k,t)$ has an elliptical form with axes along $k_x$ and $k_y$.
The white spots reveal that four peaks  with the same height
are present. 
In the plane $k_z$  a behaviour reminiscent of the two-dimensional
case is observed. 
These results can be compared and better describe the experimental measures
of the structure factor of \cite{Bey}.
As in $d=2$, the periodic modulation of the relative
heights of the different peaks produces an oscillatory behavior of most
observables, as shown in Fig.~(\ref{3d_3}) for the characteristic sizes
$R_x,R_y,R_z$ and in Fig.~(\ref{3d_4}) for the excess viscosity 
$\Delta \eta$.  
While the oscillations in the $x$ and $y$ directions
are  out of phase, due to the recurrent prevalence of thin and
thick domains described in the two-dimensional system, it is interesting
to note that $R_z$ has a different phase with respect to $R_y$,
indicating an even reacher behavior. 
These results provide at least an indication of a power-law increase
of $R_x$ and $R_z$, apart from the oscillations,  with 
an exponent larger than 1.2 for  $R_x$. 
In the $y$ direction, however, there is not even any evidence of
growth.

\section{Conclusions} \label{conc}

In this article we have considered the phase separation process
of a fluid binary mixture in the presence of an applied plane
shear flow. The binary mixture is  described by a continuum free-energy
functional and we have studied  numerically the corresponding
 time-dependent Ginzburg-Landau equation with a convective
term introduced by the flow. Results have been presented for
two-dimensional systems quenched to different final temperatures 
and for three-dimensional fluids quenched to $T=0$.
Simulations were carried out on lattices larger than those usually
considered in phase separation studies in order to get  the highest
possible resolution 
for the structure factor.

In real binary fluids the
existence of a four-peaked structure factor has been reported
\cite{migler} together with a {\it double overshoot} 
of the excess viscosity. To our knowledge, however, a coherent
picture indicating a possible connection between these apparently
different phenomena with a clear physical interpretation of
their origin was not available. 
The results of this paper strongly
suggest that the interplay between the four  peaks of the structure
factor, with their recurrent prevalence, gives rise to an oscillatory
phenomenon which reflects itself on the main observables. 
On these
bases we conjecture that the {\it double overshoot} observed in experiment
can be interpreted as the first part of an oscillatory pattern
superimposed on the global trend of the excess viscosity and that 
other local maxima of this quantity could be observed in an experiment on 
longer time scales. 

Our simulations suggest that 
the physical origin of the oscillating behavior is related
to  the
presence of two kind of domains, characterized by different sizes.
The kinetics proceeds by stretching the thicker and breaking
the thinner ones , so that a cyclical prevalence of one of the two types
occurs. This is reflected on the properties of the structure factor
and other quantities. These results agree at a semi-quantitative level
with the pattern observed in the large-$N$ approximation \cite{noi},
indicating that the global picture put forward by the $N=\infty $
theory is reliable, despite the different physical character of the
large-$N$ system without stable topological defects. This predictiveness
can be used to infer that the observed oscillations are periodic
in the logarithm of time, and that the phase separation proceeds
via dynamical scaling, so that the characteristic sizes of the domains
grow according to power-laws whose expected exponents have been computed
in \cite{brayshear,prl2}. However these conjectures could not be clearly
addressed by our simulations with the available CPU resources, because
much larger system should be considered on much longer times.

\acknowledgments
F.C. is grateful to M.Cirillo, R.Del Sole and M.Palummo for hospitality 
in the University of Rome.
F.C. and G.G. acknowledge support by the TMR network contract ERBFMRXCT980183
and  by PRA-HOP 1999 INFM.

\newpage
\begin{center}
{\large FIGURES}
\end{center}

\begin{figure}
\caption{Configurations of a portion of $512\times512$ sites of the whole
lattice are shown at different values of the strain $\gamma t$ at $T=0$.
The $x$ axes is in the horizontal direction.}
\label{T0_1}
\end{figure}

\begin{figure}
\caption{The structure factor is shown at different values of the shear
strain $\gamma t$ at $T=0$. The axes $k_x$ and $k_y$ are on the horizontal
and vertical direction, respectively.}
\label{T0_2}
\end{figure}

\begin{figure}
\caption{The same structure factor as in Fig.~(\ref{T0_2}) 
is shown in a three-dimensional plot 
at $\gamma t=11$ and $\gamma t=20$ in order to better illustrate the different
relative heights of the peaks at different times. 
The axes $k_x$ and $k_y$ are on the horizontal
and vertical direction, respectively.}
\label{T0_3}
\end{figure}

\begin{figure}
\caption{The average size of domains in the flow (upper curve) and in the shear
direction (lower curve) are plotted as a 
function of the shear strain $\gamma t$ 
at different temperatures. The straight lines have slope 4/3 at 
$T=0,1,3$ and $1/3$ at $T=5$.}
\label{T0_4}
\end{figure}

\begin{figure}
\caption{The excess viscosity is plotted as a function of the shear 
strain $\gamma t$ at different temperatures. The straight lines have slope 
-5/3 at $T=0,1,3$ and $-2/3$ at $T=5$.}
\label{T0_5}
\end{figure}

\begin{figure}
\caption{Configurations of a portion of $256\times256$ sites of the whole
lattice are shown at different values of the strain $\gamma t$ at $T=5$.
The $x$ axes is in the horizontal direction. Grey-scaling from black  
to  white
corresponds to values of  $\varphi$ from -1 to 1.}
\label{Tf_1}
\end{figure}

\begin{figure}
\caption{The structure factor is shown at different values of the shear
strain $\gamma t$ at $T=5$. The axes $k_x$ and $k_y$ are on the horizontal
and vertical direction, respectively.}
\label{Tf_2}
\end{figure}

\begin{figure}
\caption{The interfaces between domains of different compositions of a 
$128^3$ section of the simulated system are shown at $T=0$.
The $x$-axis points from left to right, the $y$-axis points into the foil
and the $z$-axis from bottom to top.}
\label{3d_1}
\end{figure}

\begin{figure}
\caption{The sections $k_x=0$, $k_y=0$ and $k_z=0$ of the
structure factor are shown at $\gamma t=1, 3, 1$,  
respectively, and $T=0$.}
\label{3d_2}
\end{figure}

\begin{figure}
\caption{Evolution of the average domain size in the $x$ 
($\triangle$), 
$y$ ($\circ$) and $z$ direction ($\bullet $) at $T=0$. 
The straight line has slope $4/3$.}
\label{3d_3}
\end{figure}

\begin{figure}
\caption{The excess viscosity is plotted as function of the shear 
strain $\gamma t$ at $T=0$.
The straight line has slope $-5/3$.}
\label{3d_4}
\end{figure}


\begin{references}
\bibitem{notarella}
The characteristic lengths $R(t)$ are measured in units of
$\sqrt{\kappa/a}$ and time in terms of $\gamma a^2/\kappa$.
\end{references}
\end{document}